\documentclass[reprint,superscriptaddress,amsmath,amssymb,aps,practice]{revtex4-1}
\usepackage{graphicx}
\usepackage{amssymb}
\usepackage{amsmath}
\usepackage{color}
\usepackage{slashed}
\usepackage{ulem}

\usepackage{amsmath,amsfonts,amssymb}

\newcommand{\be}{\begin{equation}}
\newcommand{\ee}{\end{equation}}
\newcommand{\bea}{\begin{eqnarray}}
\newcommand{\eea}{\end{eqnarray}}

\makeatletter
\let\Hy@backout\@gobble
\makeatother

\begin{document}
	\title{Exact relaxation dynamics and quantum information scrambling in multiply quenched harmonic chains }
	
	\author{Supriyo Ghosh}
	\email{supriyo.physics@gmail.com}
	\author{Kumar S. Gupta}
	\email{kumars.gupta@saha.ac.in}
	\affiliation{Theory Division, Saha Institute of Nuclear Physics, HBNI, 1/AF Bidhannagar, Kolkata 700064, India}
	\author{Shashi C. L. Srivastava}
	\email{shashi@vecc.gov.in}
	\affiliation{Variable Energy Cyclotron Centre, HBNI, 1/AF Bidhannagar, Kolkata 700064, India}

	\date{\today}
	
	\begin{abstract}
		
		The quantum dynamics of isolated systems under quench 
		condition exhibits a variety of interesting features 
		depending on the integrable/chaotic nature of system. We study the \textit{exact} dynamics of trivially integrable 
		system of harmonic chains under a multiple quench protocol. Out 
		of time ordered correlator of two Hermitian operators at large time displays 
		scrambling in the thermodynamic limit. In this limit, 
		the entanglement entropy and the central component of 
		momentum 
		distribution both saturate to a steady state value. We also show that 
		reduced density matrix assumes the diagonal form long 
		after multiple quenches for large system size. These exact 
		results involving infinite dimensional Hilbert 
		space are indicative of dynamical equilibration for a 
		trivially integrable harmonic chain.
	\end{abstract}

	\maketitle

	\section{Introduction}
	The behaviour of isolated quantum systems under non-equilibrium conditions is of great interest from both theoretical \cite{rigol-gge,rigol,krishnendu,srednicky-eth,rigol-review,Deutsch_2018} as well as experimental point of view \cite{raj,mom-expt1,kaufman}. A generic method to reach a non-equilibrium regime is by a single or multiple quenches of the system parameters. For   non-integrable systems, the quench protocol often leads to thermalization  \cite{srednicky-eth,rigol}, although there can be subtleties depending on the choice of initial states \cite{calabrese-2012-26,calabrese-2012-27}.
	
	For integrable systems, the situation is not as unequivocal. It is generally known that an isolated integrable system does not thermalize \cite{rigol,Deutsch_2018}, but can be described by a generalized Gibbs ensemble (GGE) \cite{rigol-gge,rigol-gge-review,Essler_2016,mussardo-gge,wright-prl}.However, it has recently been found that the $XXX$ Heisenberg spin chain, integrable under the Bethe ansatz scheme, can exhibit weak eigenstate thermalization for a typical, but not necessarily all eigenstates \cite{alba,alba-calabrese,inte-themalization-grisins}. It is also known that certain finite dimensional systems exhibit comparable statistical relaxation regardless of whether they are integrable or not \cite{santos-prl}. Furthermore,  an integrable Jaynes-Cummings model can show rapid decoherence analogous to chaotic dynamics and the  inability to recover the purity  at large times increases in  the question of dynamical relaxation thermodynamic limit \cite{integrable-decoherence}.  For an integrable system with infinite dimensional Hilbert space, Bogoliubov and Krylov showed that under certain assumption about the thermal reservoir, the system relaxes to thermal state in thermodynamic limit (see \cite{Chirikov1986} and references therein). In 1d-bosonic systems, a class of special initial states thermalizes under integrable dynamics governed by Gross-Pitaevskii equation \cite{inte-themalization-grisins}.

	Various physical quantities have been used to analyze the onset of statistical relaxation following a quench. In a quantum many-body system, the reduced density matrix (RDM) carries important information regarding the relaxation dynamics \cite{Peschel_1999,ingo2,ingo-2004,ingo-2009,peschel-2003,essler-prb,calabrese-2012}.   The entanglement entropy calculated from the RDM measures the loss of information and its dynamics describes how the quantum information is spread.   The one body momentum distribution, also obtained from the RDM, carries the signature of the { dynamical relaxation}. 
	
	Recently, the out of time order correlator (OTOC) \cite{larkin} has gained prominence in the context of scrambling of quantum information in non-equilibrium systems \cite{maldacena2016,ganeshan-prl,swingle-nature,pollman-otoc,lin1,lin2,otoc-expt-prx,propsen-weak-q-chaos,lev-info-prop,cotler-otoc}. Although information scrambling is usually a property of chaotic systems, the OTOC of certain non-local operators exhibit scrambling even in an integrable Ising chain \cite{lin1}.  It has been further argued that scrambling could be independent of the integrability of the Hamiltonian \cite{scrambling-integrable} and that mutual information in an integrable spin chain can exhibit weak scrambling \cite{alba-calabrese-2019}. The above list of examples, which is by no means exhaustive, indicates that there are still many open questions in the context of dynamical relaxation and information scrambling in integrable systems.

	{In this paper we discuss two different dynamical aspects of an integrable harmonic chain. The first point is to address the  question of dynamical relaxation to a steady state under a multiple quench protocol. The second point is to analyze the dynamics of information scrambling under a similar set of quenches}.

	The main differences of our analysis with the existing approaches in the literature are 
	\begin{itemize}
	\item We consider a series of multiple global quenches with no restriction to their number, in contrast to a single quench which is typical in the existing literature. As we shall show, the multiple quench protocol leads to important differences  compared to a single quench.
	\item Our analysis involves systems  with infinite dimensional Hilbert space at each site and local Hermitian operators. This is particularly relevant for information scrambling and OTOC where majority of the existing analyses involves finite spin systems and unitary operators.
	\item The time development is treated in an exact analytical fashion by solving TDSE using non-linear Ermakov equation. The total number of particles and the number of quenches can be arbitrarily large and the analysis is valid for indefinitely large time.
	\end{itemize}
	
	In order to achieve the above, we study the exact relaxation dynamics following  multiple global quenches in a harmonic chain with $N$ oscillators, which is trivially integrable. For this system, the exact entanglement dynamics following a single quench has been studied earlier \cite{ghosh_2017}. Under a multiple quench protocol, we show that the off-diagonal terms in the one-particle RDM \cite{Peschel_1999, peschel-2003} vanish exponentially fast in time in the limit of large $N$, while the diagonal terms saturate to a steady value.  The mixing of a large number of incommensurate and irrational normal mode frequencies is responsible for this feature. In contrast, for a single quench, the relaxation to the steady state is much slower. Under the same conditions, the momentum distribution and the  entanglement entropy also show the signatures of a steady state.

	Universality of the small time behaviour of OTOC in chaotic systems and its relation with Lyapunov exponents has been the focus of most of the studies \cite{ganeshan-prl,maldacena2016,swingle-nature, arul-2019}. Here we want to focus on long time behaviour much after the completion of the quench protocol. The multiple quench protocol has important consequences in the quantum information scrambling and OTOC. While recurrences are characteristic feature of integrability in OTOC, in contrast for this system, OTOC saturates to a nonzero value under multiple quenches in thermodynamic limit, which is indicative of scrambling \cite{alba-calabrese-2019}. As our exact analysis is valid for arbitrarily large number of particles, we can clearly demonstrate the finite size effect by varying the number of oscillators.   
	
	The harmonic chain under consideration can be experimentally realized in various systems such as in optical tweezers \cite{science} and the tuning of individual coupling parameters can be done using ultracold atoms \cite{hunger} or Rydberg states \cite{buchmann-2017,macri}. In particular, it can be simulated using Bose-Hubbard model in the strong superfluid phase\cite{eisert-2008}. Various quench protocols have already been experimentally realized for the Bose-Hubbard model using cold atoms in optical lattices \cite{mom-expt1,kaufman,raj}. It is thus plausible that the predictions of our work can be empirically verified.
	
	This paper is organized as follow. In Sec. 2, we setup the formalism for multiple quench protocol and obtain the solutions of TDSE using non-linear Ermakov equations. In Sec. 3, we discuss the time dependence of RDM and show that the off-diagonal elements vanish for large time and in the thermodynamic limit. In Sec. 4, we obtain the momentum distribution and entanglement entropy using the RDM and discuss their steady state properties. Quantum information scrambling is discussed in Sec. 5 and it is shown that OTOC saturates to a non-zero steady value for the case of multiple quenches in the same limit. We conclude the paper in Sec. 6 with a summary of results and remarks.
	
	
	\section{Harmonic chain and the quench protocol}
	
	We consider an isolated harmonic oscillator chain with $N$ oscillators and with periodic boundary condition. The Hamiltonian is given by
	\begin{equation}
	\label{H}
	\begin{aligned}
	H(t)&=\frac{1}{2}\left[\sum_{j=1}^N(p_j^2+\omega^2(t)x_j^2)+k(t)\sum_{j=1}^{N}(x_j-x_{j+1})^2\right]\\
	\\
	&=\frac{1}{2}\left[\sum_{j=1}^Np_j^2+X^T.\Sigma(t).X\right],
	\end{aligned}
	\end{equation}
	where $X$=$(x_1,x_2,...,x_N)^T$ and $\Sigma$ is an $N\times N$ real symmetric matrix. Here the frequency $\omega$ and the nearest neighbour coupling $k$ are explicit functions of time. A solution of the corresponding time dependent Schr\"odinger's equation (TDSE) \cite{osc1,osc2,ghosh_2017}
	\begin{equation}\label{tdse}
	i\frac{\partial}{\partial t}\mid\psi\rangle=H(t)\mid\psi\rangle
	\end{equation}
	can be written as
	\begin{equation}\label{wf}
	\begin{aligned}
	\psi(x_1,....,x_N,t)=&\left(\text{det} \frac{\Omega}{\pi}\right)^{\frac{1}{4}}\exp\left[i\left(X^T\tilde{b}X-\sum_{j=1}^NE_j\tau_j\right)\right]
	\\
	&~~\times\exp\left[-\frac{X^T\Omega X}{2}\right],
	\end{aligned}
	\end{equation}
	where $\Omega=U^T\sqrt{\Sigma^{\prime D}}U$, $\Sigma_{jj}^{\prime D}=\frac{\Sigma_{jj}^D(0)}{b_j^4(t)}$, $\tilde{b}=U^T\tilde{b}^DU$,$\tau_j=\int_0^t\frac{dt^\prime}{b_j^2(t^\prime)}$, $U$  is an  orthogonal transformation which transforms the matrix $\Sigma$ to its diagonalized form $\Sigma^D$ and $\tilde{b}^D$ is a diagonal matrix with elements $\frac{\dot b_j(t)}{2b_j(t)}$. Here $b_j(t)$ satisfies the the nonlinear Ermakov equation \cite{osc1,osc2,pinney} given by    
	\begin{equation}
	\label{4-erm}
	\ddot b_j+\lambda_j(t)b_j=\frac{\lambda_j(0)}{b_j^3},
	\end{equation}
	where $\lambda_j$'s are the normal mode frequencies of the Hamiltonian, which have the form
	\begin{equation}\label{lmda}
	\lambda_j(t)=\omega^2(t)+2k(t)-2k(t)\cos(2\pi j/N)
	\end{equation}
	with $\lambda_j = \lambda_{N-j}$.
	 We choose the initial condition as the ground state of the time independent pre-quenched Hamiltonian of the $N$ oscillator chain, which requires that $b(0) =1$ and $\dot{b}(0) = 0$. Note that  finding the solutions of the TDSE is equivalent to finding the solutions of the Ermakov equations (\ref{4-erm}). The normalization is determined by the condition that $ \int_{-\infty}^\infty \psi^*(x,t) \psi(x,t) dx = 1 $ \cite{osc2}.

	\textbf{\textit{Solution for series of quenches:}}
	Our quench protocol is shown in Fig. (1). At time $t=0$, the 
	frequency $\omega(i)$ is quenched to $\omega(f)$ and $k(=2)$ is 
	left unchanged. This defines a single quench. After a time $T$, 
	the frequency $\omega(f)$ is changed back to $\omega(i)$, 
	defining the second quench. This sequence is repeated till the 
	required number of quenches is achieved. Now we shall discuss the 
	solution of Ermakov equations under $n$ such quenches. 
In terms of the variable 
	\begin{equation}\label{def-eta}
	\eta_j(t) \equiv b_j^2(t),
	\end{equation}
	 the Ermakov equations (\ref{4-erm}) can be written as 
	\begin{equation}
	\label{forB}
	\ddot \eta_j \eta_j-\frac{1}{2} 
	\dot\eta_j^2+2\lambda_j(t)\eta_j^2=2\lambda_j(0).
	\end{equation}
	\begin{figure}
		\label{fig1}
		\centering
		\includegraphics[width=8.6cm]{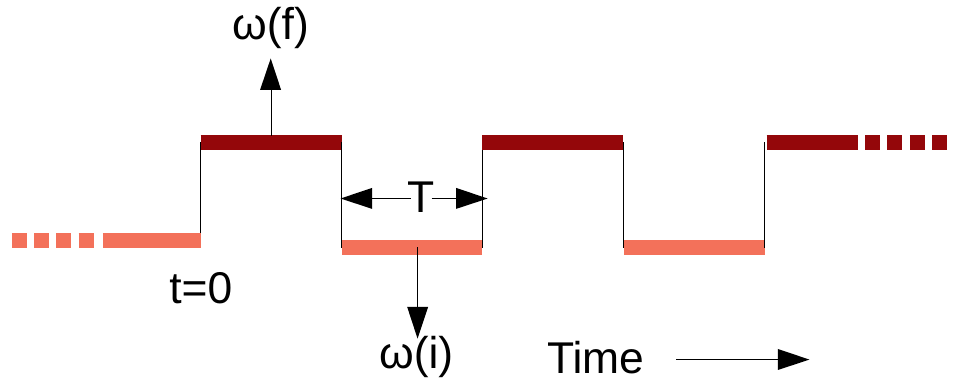}
		\caption{The schematic diagram of $\omega(t)$ for multiple quench protocol. For all our calculations, we quench from $\omega(i)=3$ to
		$\omega(f)=20$ and time interval between two quenches is taken as 4 seconds.}
	\end{figure}
	 The corresponding solutions are given by 
	\begin{align}\label{soln}
	&\nonumber\eta_{j,1}(t)=\alpha_{j,1} \cos(2\sqrt{\lambda_{j}(t)} t)+\beta_{j,1} \sin(2\sqrt{\lambda_{j}(t)} t)+\gamma_{j,1}\\\nonumber&~~~~~~~~~~~~~~~~~~~~~~~~~~~~~~~~~~~~~~~~~~~~~~~~~~~~~~ 0<t<T\\
	&\nonumber\eta_{j,2}(t)=\alpha_{j,2} \cos(2\sqrt{\lambda_{j}(t)} t)+\beta_{j,2} \sin(2\sqrt{\lambda_{j}(t)}t)+\gamma_{j,2}\\\nonumber&~~~~~~~~~~~~~~~~~~~~~~~~~~~~~~~~~~~~~~~~~~~~~~~~~~~~~~T<t<2T\\\nonumber
	&\eta_{j,i}(t)=\alpha_{j,i} \cos2(\sqrt{\lambda_{j}(t)} t)+\beta_{j,i} \sin(2\sqrt{\lambda_{j}(t)}t)+\gamma_{j,i}\\\nonumber&~~~~~~~~~~~~~~~~~~~~~~~~~~~~~~~~~~~~~~~~~~~~~~~~~~(i-1)T<t<iT\\\nonumber & \vdots  \\\nonumber & \eta_{j,n}(t)=\alpha_{j,n} \cos2(\sqrt{\lambda_{j}(t)} t)+\beta_{j,n} \sin(2\sqrt{\lambda_{j}(t)}t)+\gamma_{j,n}\\&~~~~~~~~~~~~~~~~~~~~~~~~~~~~~~~~~~~~~~~~~~~~~~~~(n-1)T<t<\infty
	\end{align}
	where $\alpha_{j,i}$, $\beta_{j,i}$ and $\gamma_{j,i}$ are the time independent constants which are obtained from the boundary conditions on $b$ or $\eta$. Furthermore, using the above equation we could express $\eta_j$ and its derivatives in a matrix form as
	\begin{equation}
	\begin{aligned}
	\vec\eta_j(t)=
	\begin{pmatrix} 
	\eta_j(t) \\ \dot\eta_j(t) \\ \ddot\eta_j(t)  
	\end{pmatrix}=A_j(t)\begin{pmatrix} 
	\alpha_{j,i} \\ \beta_{j,i} \\ \gamma_{j,i}  
	\end{pmatrix},
	\end{aligned}
	\end{equation}
	where 
	{\small
		\begin{equation}\label{A-matrix}
		\begin{aligned}
		&A_j(t)=\\& 
		\begin{pmatrix} 
		\cos(2\sqrt{\lambda_{j}(t)}t) & \sin(2\sqrt{\lambda_{j}(t)}t) &  1 \\  
		-2\sqrt{\lambda_{j}(t)}\sin(2\sqrt{\lambda_{j}(t)} t) & 
		2\sqrt{\lambda_{j}(t)}\cos(2\sqrt{\lambda_{j}(t)} t) &  0 \\ -4\lambda_{j}(t) \cos(2\sqrt{\lambda_{j}(t)}t) & 
		-4\lambda_{j}(t)\sin(2\sqrt{\lambda_{j}(t)} t) &  0  
		\end{pmatrix} .
		\end{aligned}
		\end{equation}
	}
	The continuity of the wavefunction across the quenches implies the continuity in 
	 $\eta_j(t)$ and $\dot\eta_j(t)$. $\vec \eta_j(t^+)$ and $\vec \eta_j(t^-)$ defined just after and before any quench respectively are related by	
	\begin{equation}
	\begin{aligned}
	\vec\eta_j(t^+)=
	B_j(t^+\leftarrow t^-)\vec\eta_j(t^-) ,
	\end{aligned}
	\end{equation}
	where the form of B matrix could be found  using Eq.(\ref{forB}) as
	\begin{equation}
	\begin{aligned}\label{B-matrix}
	B_j=
	\begin{pmatrix} 
	1 & 0 &  0 \\  0 & 1 &  0 \\ 2(\lambda_{j}(t^-)-\lambda_{j}(t^+)) &0 &  1  
	\end{pmatrix}.
	\end{aligned}
	\end{equation}
	 The coefficient vectors with elements $(\alpha, \beta, \gamma)_j$ across the quench are related as
	\begin{equation}\label{coeff}
	\begin{aligned}
	\begin{pmatrix} 
	\alpha_{j,i+1 }\\ \beta_{j,i+1} \\ \gamma_{j,i+1}  
	\end{pmatrix}
	=A^{-1}(t^+)B(t^+\leftarrow t^-)A(t^-)\begin{pmatrix} 
	\alpha_{j,i} \\ \beta_{j,i} \\ \gamma_{j,i}  
	\end{pmatrix},
	\end{aligned}
	\end{equation}
	with the boundary condition,
	\begin{equation}\label{BC}
	\begin{aligned}
	\begin{pmatrix} 
	\alpha_{j,1} \\ \beta_{j,1} \\ \gamma_{j,1}  
	\end{pmatrix} =
	\begin{pmatrix} 
	\frac{\lambda_{j,f} - \lambda_{j,in}}{2\lambda_{j,f}}  \\
	0 \\   \frac{\lambda_{j,f} + \lambda_{j,in}}{2\lambda_{j,f}}
	\end{pmatrix} 
	\end{aligned}.
	\end{equation}
	This gives a complete solution of the Ermakov equations for the given quench protocol.
	
	
	\section{Reduced density matrix}
	
		In this Section we shall discuss the time-dependent one-body RDM obtained under a series of quenches in the coordinate representation. This RDM would subsequently be used to obtain the one-body momentum distribution and the entanglement entropy.

In order to obtain the time dependent one-body RDM, we trace out the rest of the system except the $i^\text{th}$ site. The RDM has the form
	\begin{equation}\label{den1}
	\begin{aligned}
	\rho_{red}(x_i,x_i^\prime,t)=\int \prod_{a=1\neq i}^N dX^{a}dX^{\prime a} \psi^*(X^a,t)\psi(X^{\prime a},t).
	\end{aligned}
	\end{equation}
	Using (\ref{wf}) in (\ref{den1}) and carrying out the integration, we get  
	\begin{equation}\label{RD1}
	\begin{aligned}
	\rho_{red}(x_i,x_i^{\prime},t)=& \left(\frac{\gamma-\beta}{\pi}\right)^{\frac{1}{2}}\exp\left[iZ\left (x_i^2-x_i^{\prime 2}\right )\right]\\&\times\exp\left[-\frac{1}{2}\gamma\left(x_i^{2}+x_i^{\prime 2}\right)+x_i\beta x_i^{\prime }\right],
	\end{aligned}
	\end{equation}
	where $Z$, $\beta$, $\gamma$ are given by
	\begin{equation}
	\begin{aligned}\label{beta-gamma}
	&Z(t)=\tilde{b}_{1 \times 1}-\tilde{b}_{N-1 \times 1}^{\dagger}\Omega_{N-1\times N-1}^{-1}\Omega_{N-1\times 1},\\
	&\gamma(t)=\Omega_{1\times 1}-\frac{1}{2}\Omega_{N-1\times 1}^{\dagger}\Omega_{N-1\times N-1}^{-1}\Omega_{N-1\times 1}\\
	&~~~~~~~~+2\tilde{b}_{N-1 \times 1}^{\dagger}\Omega_{N-1\times N-1}^{-1}\tilde{b}_{N-1 \times 1}, \\
	&\beta(t)=\frac{1}{2}\Omega_{N-1\times 1}^{\dagger}\Omega_{N-1\times N-1}^{-1}\Omega_{N-1\times 1}\\&~~~~~~~~+2\tilde{b}_{N-1 \times 1}^{\dagger}\Omega_{N-1\times N-1}^{-1}\tilde{b}_{N-1 \times 1}.
	\end{aligned}
	\end{equation}
	
	A generic off-diagonal element of the RDM has the form 
	\begin{equation}\label{RD2}
	\begin{aligned}
	&\rho_{red}(x_i,x_i+\Delta,t)= \left(\frac{\gamma-\beta}{\pi}\right)^{\frac{1}{2}}\exp\left[-iZ\left (\Delta x_i+\Delta^2\right )\right]\\&\times\exp\left[-(\gamma-\beta)\left(x_i-\frac{\Delta}{2}\right)^2-\Delta^2\left(\gamma-\frac{\gamma-\beta}{4}\right)\right],
	\end{aligned}
	\end{equation}
	where $\Delta$ denotes the distance from the diagonal axis of the RDM and $\Delta=0$ corresponds to the diagonal elements of the RDM. The equation (\ref{RD2}) shows that the off diagonal elements are shifted Gaussians with a time dependent exponent. The ratio $r$ between the diagonal and a generic off-diagonal element can be written as
	\begin{align}\label{ratio}
	\nonumber r&=\frac{\rho_{red}(x_i,x_i+\Delta,t)}{\rho_{red}(x_i,x_i,t)}\\&=\exp\left[-iZ\left (\Delta x_i+\Delta^2\right)\right]\exp\left[-(\gamma-\beta)x_i\Delta-\gamma\Delta^2\right].
	\end{align}

	\begin{figure}
		\centering
			\includegraphics[width=7cm]{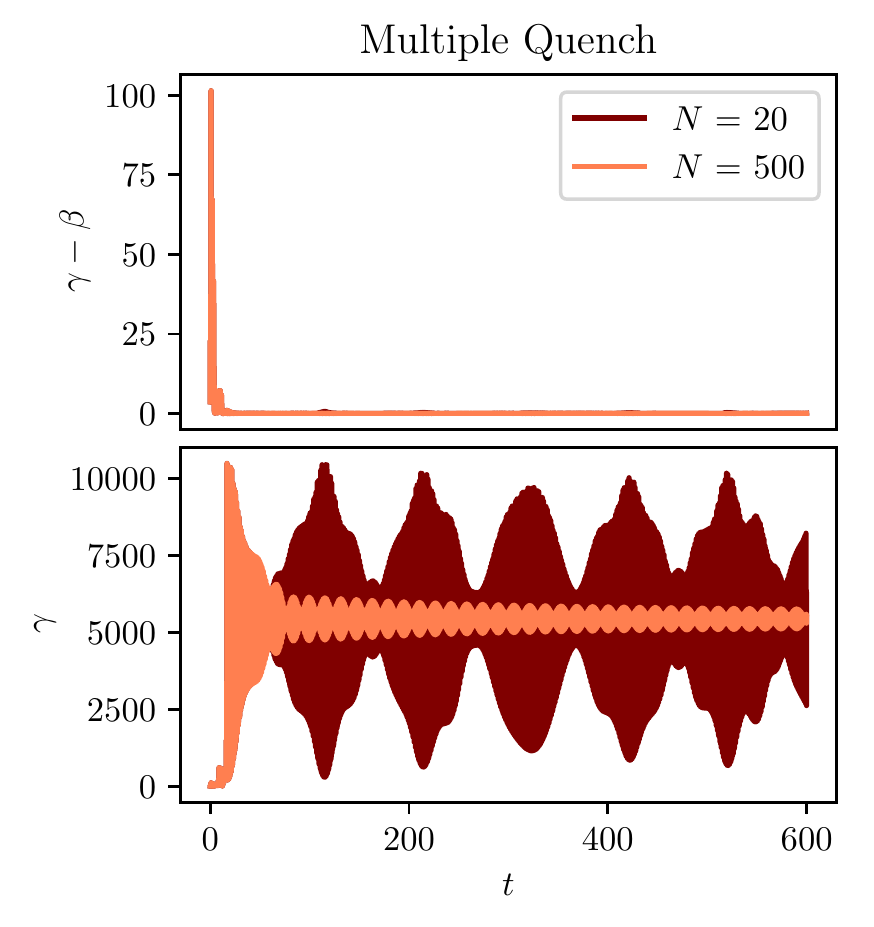}
		\includegraphics[width=7cm]{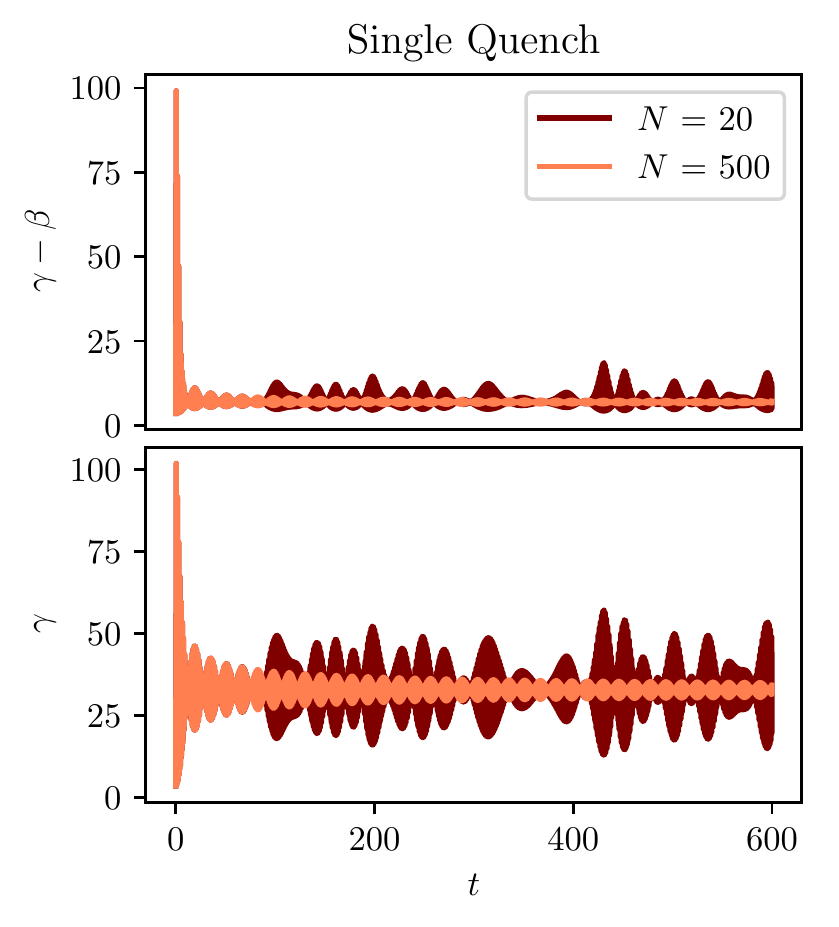}
			\caption{Plot of $\gamma$ and $(\gamma-\beta)$ as a function of time, for  multiple (five) quenches and a single quench. An order of magnitude higher value of $ \gamma$ in case of multiple quenches for large $ N $ brings the RDM to the diagonal form exponentially faster as compared to a single quench. For smaller values of $ N $, $ \gamma $ and correspondingly off-diagonal elements of RDM oscillate in time. The $ (\gamma - \beta) $ is of the same order for both protocols but saturates near zero value in the multiple quench scenario in contrast to approximately $ 8 $ for single quench.}
\label{fig:2}
	\end{figure}
	
	The exponents $\gamma$ and $(\gamma-\beta)$ are plotted in Fig. \ref{fig:2} and are found to be real positive for all time. This also follows from the reality of the eigenvalues of the RDM, which is Hermitian \cite{ghosh_2017}.  For a series of multiple quenches we find that $(\gamma - \beta)$ rapidly saturates to a small positive value  irrespective of $N$. However $\gamma$ has very different behaviour depending on the value of $N$. For higher $N$, $\gamma$ has a large value which oscillates very little with time. For smaller $N$, the mean value of $\gamma$ is also high but the fluctuations are very big as well. On the other hand, for a single quench, both $\gamma$ and $(\gamma-\beta)$ show appreciable oscillations and the value of $\gamma$ is much less irrespective of the value of $N$. 
	
	In Eq.(\ref{ratio}) we have given the ratio of the off-diagonal to the diagonal elements of the RDM. We thus find that for multiple quenches and in the thermodynamic limit, the off-diagonal matrix elements of the RDM tend to zero. For a single quench, the suppression of the off-diagonal matrix elements with time is much less.

	Hence, for multiple quenches with the large system size, the RDM assumes a diagonal form at a time large compared to the duration of the quenches. This can be qualitatively understood as follows. The solution of the Ermakov equation for each normal mode contains an irrational number given by the square root of corresponding normal mode frequency $\lambda_j$. As the number of oscillators increases, a large number of irrational and incommensurate frequencies start contributing to the wave function and the RDM. The mixing of a large number of such modes is responsible for the statistical relaxation of the quantities $\gamma$ and $(\gamma - \beta)$ with time, which in turn reduces the RDM to a diagonal form.


	\section{momentum distribution and entropy}
	
	We derive the analytical expression of one body momentum distribution by taking the Fourier transform of (\ref{RD1}), which is given by
	\begin{align}
	\nonumber n(p,t)&=\frac{1}{2\pi}\int dx_idx_i^{\prime}\rho_{red}(x_i,x_i^{\prime},t)e^{-ip(x_i-x_i^{\prime})}\\&=\int dx_idx_i^{\prime} \left(\frac{\gamma-\beta}{4\pi^3}\right)^{\frac{1}{2}} e^{[i Z(x_i^2-x_i^{\prime 2})-ip(x_i-x_i^{\prime})]}\\\nonumber&~~~~~\times e^{[-\frac{\gamma}{2}(x_i^2+x_i^{\prime 2})+\beta x_ix_i^{\prime}]}.
	\end{align}
	After the integration, the momentum distribution takes the form
	\begin{align}\label{mom}
	n(p,t)=\left(\frac{\gamma-\beta}{\pi\left(\gamma^2-\beta^2+4Z^2\right)}\right)^{\frac{1}{2}} e^{\left[\frac{-p^2(\gamma-\beta)}{\gamma^2-\beta^2+4Z^2}\right]}.
	\end{align}
	
\begin{figure}
	
	\centering
	\includegraphics{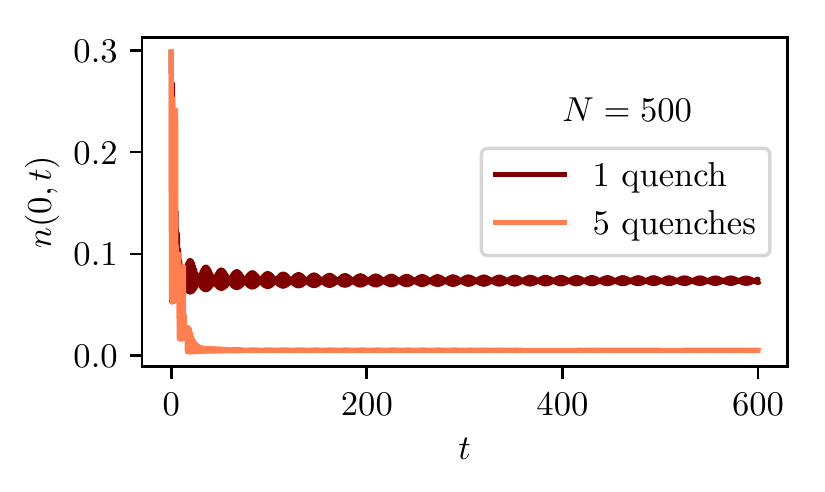}
	\caption{The central component of momentum distribution, $n(0,t)$ is plotted as a function of time for single and  multiple quenches. The fluctuation decreases substantially with increasing number of quenches.}\label{fig3}
\end{figure}

	At time $t=0$ just before the quench, the distribution (\ref{mom}) is a Gaussian with mean at $p=0$.  In Fig. \ref{fig3} the central component of the momentum distribution is plotted with time for both single and multiple quenches and with $N=500$. For a single quench, in the long time limit the momentum distributions reaches a steady value but still shows appreciable oscillations.  For multiple quenches the value in the long time limit is lower and the fluctuations are negligible. Thus in the long time and in the thermodynamic limit, the higher the number of quenches, the better is the relaxation of the system to a steady state. 
	
	%

	In order to study the entanglement entropy, we bipartite the system into two parts of one oscillator versus $N-1$ oscillators. The entanglement entropy of the smaller subsystem calculated using the reduced density matrix (\ref{RD1}) has the form \cite{ghosh_2017}
	\begin{align}
	\label{vn}
	S(t)=-\log(1-\xi(t))-\frac{\xi(t)}{1-\xi(t)}\log\xi(t),
	\end{align}
	where the $\xi$ has the form given by
	
	\begin{align}\label{eq:xi}
	\xi(t)=\frac{\frac{\beta}{\gamma}}{1+\sqrt{1-\frac{\beta^2}{\gamma^2}}} < 1,
	\end{align}
	$\beta $, $\gamma$ are given in eq. (\ref{beta-gamma}).

	\begin{figure}
		\centering
		\includegraphics{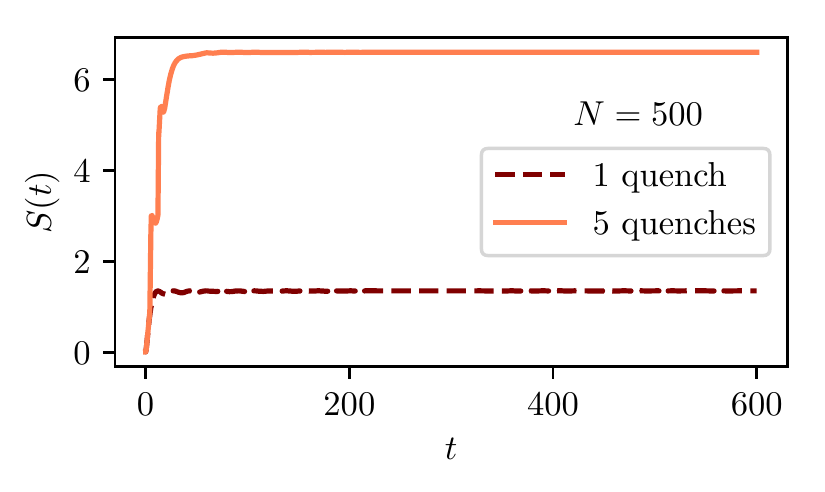}
		\caption{Evolution of entanglement entropy is plotted for single and multiple quenches. Generation of higher entanglement for multiple quenches is explained using Eq.\ref{vn} and with the fact that $ \gamma $ approaches  $ \beta $ faster as compared to a single quench.}\label{fig4}
	\end{figure}
	
	The von Neumann entropy as a function of time is plotted in Fig. \ref{fig4}. In a finite dimensional system, the maximum entropy is proportional to the Hilbert space dimension. The bound on entanglement entropy is saturated for generic chaotic systems. Here the system being infinite dimensional, there is no finite limit to the maximum entanglement entropy. As in the case of the momentum distribution,  higher number of quenches leads to smaller fluctuations in the entanglement entropy and a larger steady state value. This can be qualitatively understood in terms of reduced variation in eigenvalues of RDM given by $ p_n = (1-\xi)\xi^n, n=0,1,2,\dots $ \cite{ghosh_2017}. The variation in $ p_n $ reduces as $ \xi \to 1 $ or in other words as $ (\gamma-\beta) \to 0 $ (see Fig. \ref{fig:2}).


	\section{Quantum information scrambling}
	
The  OTOC between two operators $M$ and $N$ separated by a lattice distance $l$ is defined as      
	\begin{align}
	F(l,t)=\langle [M_i(t), N_j(0)]^{\dagger}[M_i(t), N_j(0)]\rangle,\label{otocom}
	\end{align}
where $ l = j-i $. For well separated local operators, the OTOC starts from a zero value and then increases as the information propagates with Lieb-Robinson velocity. 

\begin{figure}
	\centering
	\includegraphics{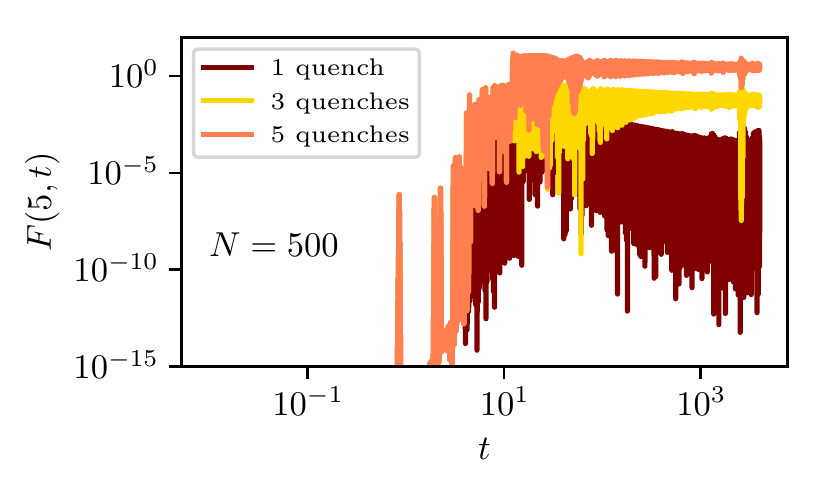}
	\caption{OTOC for Hermitian operator $ x(t) $ and $ p(0) $ between sites 1 and 6 for different number of quenches are shown. The decay for single quench rules out the scrambling. In contrast, for multiple quenches it saturates to a non-zero value which increases with number of quenches, indicative of scrambling.}	\label{fig:otoc-diff-quench}
\end{figure}	
	
 Here we consider the OTOC between position and momentum operators, which are local and Hermitian. We choose $M_i(t)=x_i(t)$, $N_i(0)=p_i(0)$ which are labeled by the site index. 
The expression for $x_i(t)$ in the Heisenberg picture for a single time dependent quantum harmonic oscillator with frequency $\omega(t)$ is given by \cite{heisenberg-ho}
	\begin{align}\label{eq:xt}
&\nonumber	x(t)=\\& x(0) b(t)\cos\left(\int_0^t\frac{\omega(0)}{b^2(t^{\prime})}dt^{\prime}\right) + p(0)\frac{b(t)}{\omega(0)}\sin\left(\int_0^t\frac{\omega(0)}{b^2(t^{\prime})}dt^{\prime}\right),
	\end{align}
where $ b(t) $ is the solution of corresponding Ermakov equation. Using Eq.(\ref{eq:xt}) with the canonical commutation relations $[x_i,p_j] = i \hbar \delta_{ij}$ we get,		
	{\small
		\begin{align}\label{otoc}
		&\nonumber F(l,t)=\langle[x_i(t), p_j(0)]^2\rangle \\\nonumber&=\left(\sum_{m=1}^NU_{mi}^\dagger b_m(t)\cos\left(\tan^{-1}\sqrt{\frac{\lambda_m(t)}{\lambda_m(0)}\tan(\sqrt{\lambda_m(t)}}\right))U_{mj}^\dagger\right)^\dagger\\&\times\left(\sum_{m=1}^NU_{mi}^\dagger b_m(t)\cos\left(\tan^{-1}\sqrt{\frac{\lambda_m(t)}{\lambda_m(0)}\tan(\sqrt{\lambda_m(t)}}\right))U_{mj}^\dagger\right).
		\end{align}
	}

\begin{figure}
	\centering
	\includegraphics{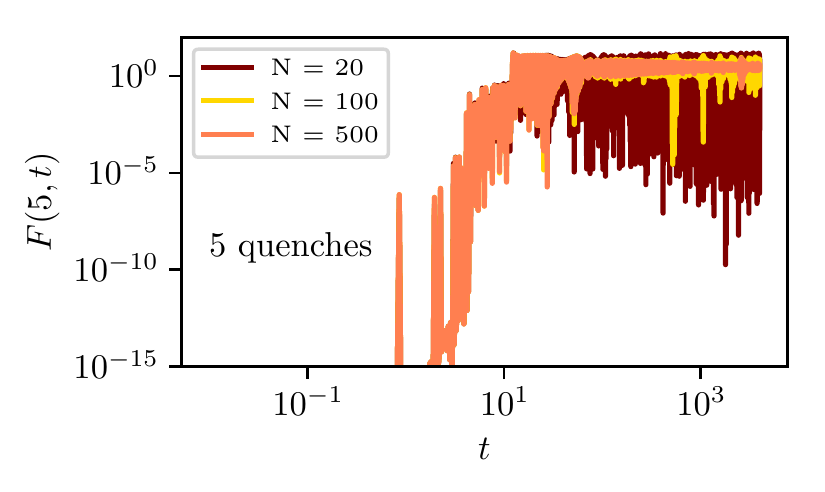}
	\caption{OTOC as a function of time is plotted for different system sizes with number of quenches fixed to 5. The finite size effects and quasi-recurrences reduce with increasing $ N $. }	\label{fig:otoc-diff-N}
\end{figure}

	In Fig.(\ref{fig:otoc-diff-quench}), OTOC is plotted for a fixed system size $N=500$ and with different number of quenches. We observe that OTOC saturates to a non-zero value with very small fluctuation for the case of multiple quenches in contrast to a constant slow decrease with large fluctuation for single quench.  A strong system size dependence on the OTOC is clearly seen from Fig. \ref{fig:otoc-diff-N}. We find that even with multiple quenches, the OTOC fluctuates enormously for smaller number of particles and does not saturate to any steady value. In Fig.(\ref{fig:otoc-diff-quench}), the quasi-recurrence in the case of multiple quenches for $N=500$ occurs at a time $t \sim \frac{N}{v_{max}}$ where $v_{max}$ is the Lieb-Robinson velocity for this system with the given set of parameters after quench.  We also note that the quasi-recurrences decrease in the thermodynamic limit as well as with the higher number of quenches.

	The saturation of the OTOC to a non-zero steady value is indicative of information scrambling. Even though the system under consideration is integrable, we find that under multiple quenches and in the thermodynamic limit, OTOC saturates to a non-zero steady value. In this formalism, the saturation can be attributed to the mixing of large number of modes with incommensurate frequencies. In the conventional picture of expanding the state as superposition of eigenfunctions of post-quench Hamiltonian, the multiple quenches would amount to accessing larger proportions of Hilbert space. This is also consistent with larger entanglement entropy with increasing number of quenches.

	
	\section{conclusion}

	In this paper we have analyzed the relaxation dynamics and quantum information scrambling in an isolated harmonic chain under multiple quenches. The various physical quantities show remarkably different non-equilibrium behaviour under the  multiple quench protocol compared to a single quench. The RDM has been shown to assume a diagonal form exponentially fast compared to a single quench. The entanglement dynamics and the momentum distribution also show relaxation to a steady state. The exact analytical results obtained here are valid for arbitrary number of particles. In order to demonstrate the finite size effects, we have graphically exhibited our results for $N=20$ and $N=500$. It is clearly seen that the quasi-revivals of the physical observables, characteristic of the finite size effects, reduce remarkably as $N$ is increased. 
	
	The conclusions are similar for the quantum information scrambling and OTOC. For a single quench and for a low value of $N=20$, the OTOC shows almost complete revival and the scrambling is practically non-existent. On the other hand, for five quenches and with $N=500$, the OTOC saturates to a finite steady value with very little revival. It may be noted that we have considered Hermitian operators to evaluate the OTOC and the system has an infinite dimensional Hilbert space. This is very different compared to the usual finite spin systems where the OTOC is normally evaluated using unitary operators. 
	
	The above result indicates that for multiple quenches and in the thermodynamic limit, the integrable harmonic chain relaxes to a steady state and the OTOC exhibits quantum information scrambling. In our formalism this has been achieved using an exact solution of the TDSE valid throughout the quench protocol and all the observables have been evaluated using the exact solution. As the number $N$ of the oscillators is increased, a large number of irrational and incommensurate normal mode frequencies start contributing to the observables whose number is of the order of $N$. The mixing of these incommensurate frequencies leads to the steady state and saturating behaviour of the various physical quantities.
	
	{ One question that has not been addressed here is what is the nature of the equilibrated state. For an integrable harmonic chain, the emergence of GGE has already been established \cite{lauchli-gge}. Whether this happens for the system under consideration with multiple quenches remains to be seen. A related question is the emergence of weak eigenstate thermalization, which has recently been observed for certain integrable systems \cite{alba,alba-calabrese,inte-themalization-grisins}. In this paper, we have focused only on a single state of the TDSE whereas to address the questions on thermalization, a more complete knowledge of the time dependent many-body spectrum would be needed. This is beyond the scope of the present analysis, which we hope to address these in future work.}

	\bibliography{rlx}

\end{document}